\documentclass[11pt]{JHEP3} 

\usepackage{epsfig}
\usepackage{graphicx}
\usepackage{amsmath}
\usepackage{yfonts}
\usepackage{bm}
\usepackage{subfigure}
\usepackage{bm}

\newcommand{\be}{\begin{equation}}
\newcommand{\ee}{\end{equation}}
\newcommand{\bea}{\begin{eqnarray}}
\newcommand{\eea}{\end{eqnarray}}
\newcommand{\ba}{\begin{array}}
\newcommand{\ea}{\end{array}}

\def\a{\alpha}

\def\o{\omega}

\def\r{\rho}

\def\t{\tau}

\def\O{\Omega}

\title{Symmetry energy of dense matter in holographic QCD }

\author{Youngman Kim
        \\
Asia Pacific Center
for Theoretical Physics and Department of Physics, Pohang University
of Science and Technology, Pohang, Gyeongbuk 790-784, Korea
\\  E-mail: \email{ykim@apctp.org}}

\author{Yunseok Seo
\\
Center for Quantum Spacetime, Sogang University, Seoul 121-742, Korea
\\ E-mail: \email{yseo@sogang.ac.kr}}

\author{Ik Jae Shin
\\
Asia Pacific Center
for Theoretical Physics, Pohang, Gyeongbuk 790-784, Korea
\\ E-mail: \email{geniean@apctp.org}}

\author{Sang-Jin Sin
\\
Department of Physics Hanyang University, Seoul 133-791, Korea
\\ E-mail: \email{sjsin@hanyang.ac.kr}}
\received{\today}       

\abstract{
We study the nuclear symmetry energy of dense matter using holographic QCD.
To this end, we consider   two flavor branes with  equal quark masses
in a D4/D6/D6 model.   We find  that
at all densities  the symmetry energy monotonically increases.
At small densities,  it exhibits a power law behavior with the density, $E_{\rm sym} \sim \rho^{1/2}$.
}


\begin{document}

\section{Introduction}

Nuclear symmetry energy is one of key words in nuclear physics as well as  in astrophysics.
 Its density dependence is a core quantity of asymmetric nuclear matter which
has important effects on heavy nuclei and is essential to understand neutron star properties.
Although much  efforts have been given,  it is still very poorly understood
especially in the  supra-saturation density regime, see \cite{sE1,sE2,sE3,sE5,sE6,sE7,Lee:2010sw} for a review and for a recent discussion.

\begin{figure}[!ht]
\begin{center}
\includegraphics[angle=0, width=0.42\textwidth]{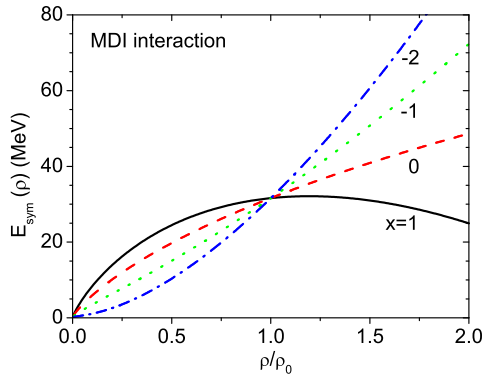}
\caption{(Color online) Example of density dependence  of the nuclear symmetry energy,  taken from \cite{Che05a}.
Depending on the value of the parameter x,  various  high density behaviors are possible.
  } \label{ty_symE}
\end{center}
\end{figure}

From experimental side, the available data
 do not constrain much  the value of the symmetry energy at supra-saturation densities.
Recently, using the FOPI data on $\pi^-/\pi^+$ ratio in central heavy ion collisions, Xiao et al.~\cite{XLCYZ} obtained a circumstantial evidence for
a soft nuclear symmetry energy at $\rho\ge 2\rho_0$, where the nuclear symmetry energy increases with the density up to
the saturation density $\rho_0$ and then starts to decrease afterwards.
Theoretically, almost all possible tools
were employed to study the density dependence of the symmetry energy. While  they showed similar  behaviors up to the nuclear saturation density, at supra-saturation  densities, all possible results one can imagine were predicted and no consensus  could be reached:  some showed
stiff dependence (increasing monotonically with density), while others showed soft one, see Fig. \ref{ty_symE} for a typical
example.  See  also \cite{sE3} for a review.
Given this situation, it would be very interesting if we can  examine the
behavior of the nuclear symmetry energy at high densities
with  a reliable calculational tool.

The gauge/gravity duality~\cite{Maldacena:1997re, Gubser:1998bc, Witten:1998qj} provides
a new  tool to study strongly interacting dense matter, and
 a few models for QCD~\cite{D4D6_03,hQCD} based on the duality were constructed.
Although the true holographic dual of  QCD is yet to be constructed,
it is worthwhile to find out what the new tool says about QCD using available models mimicking the dual of QCD.
A way to treat the  dense matter in confined phase
was suggested in \cite{Seo:2008qc}, and a model for transition from nuclear matter to strange matter  was proposed in \cite{KSS2010}.
The purpose of this paper is to calculate the symmetry energy of nuclear  matter in this model.
We will find that the symmetry energy is increasing with the total charge $Q$, showing that the symmetry energy of our system has a stiff dependence  on the density.
Also, we will explicitly calculate the density dependence of the symmetry energy at low density to show  $E_{\rm sym} \sim \rho^{1/2}$.

\section{Nuclear symmetry energy in D4/D6/D6 model}
The nuclear symmetry energy is defined as the energy per nucleon required
to change isospin symmetric nuclear matter to pure neutron matter.
In the Bethe-Weizs\"{a}cker mass formula  for the nuclear binding
energy, it represents the amount of binding energy that a nucleus has to lose when the numbers
of protons and neutrons are not equal.
The semi-empirical mass formula based on the liquid drop model has the form:
\bea
&&E_{\rm B}=a_{\rm v} A-a_a (N-Z)^2/A -a_cZ^2/A^{1/3} \nonumber \\
&&~~~~~~~~-a_sA^{2/3}\pm a_\delta/A^{3/4}\, .\label{BWf}
\eea
Here $Z$ ($N$) is the number of protons (neutrons) in a nucleus.
The first term is called the volume energy since the volume of a nucleus is proportional to $A$, where $A$ is the total nucleon number.
The origin of this volume term is the strong nuclear force.
The second is known as the asymmetry term, which defines the symmetry energy.
If there were no Coulomb repulsions between protons, we would expect to have equal number of neutrons and protons in
nuclei in general.
The term with $a_c$ accounts for the Coulomb interaction of all pairs of protons in the nucleus.
The last two terms represent the surface energy and pairing effect, respectively.
Using data for nucleus binding energies,  one can determine a set of coefficients in Eq.~(\ref{BWf}).

Due to the  invariance of nuclear
forces under neutron-proton interchange, iso-scalar quantities in a nuclear system
are function of only even powers of the asymmetry factor $\tilde\alpha$ defined by
 $\tilde\alpha\equiv (N-Z)/A$. Then we can express the energy density per nucleon $E(\rho,\tilde\alpha)$ as
\bea
E(\rho,\tilde\alpha)\simeq E(\rho,0) +S_2(\rho) \tilde\alpha^2 \, ,\label{nmE}
\eea
where $\rho$ is the nucleon number density and $S_2(\rho)=\frac{1}{2}\frac{\partial^2E}{\partial\tilde\alpha^2}\vert_{\tilde\alpha=0}$ is the symmetry energy.

Now we study the symmetry energy in the D4/D6/D6 model with  baryon vertices which consist of compact D4 branes and fundamental strings~\cite{KSS2010}.
The gluon dynamics is replaced by the gravity sourced by the $N_c$ color D4 branes,
and two probe D6 branes are used to describe the up and down quarks.
The bare quark masses are the distances between the $N_c$ color D4 and two D6's in the absence of the string coupling.

We can write the metric of the confining D4 background as
\bea\label{bgmetric}
ds^2 = \left( {U }/{R }\right)^{3/2}\left(- dt^2 +d\vec{x}^2 + f(U) dx_4^{2} \right)
+\left( {R}/{U }\right)^{3/2}\left( {U}/{\xi}\right)^2\left(d\xi^2 +\xi^2 d\Omega_4^2 \right),\nonumber
\eea
where $f(U)=1-(U_{KK}/U)^3$  and $(U/U_{KK})^{3/2}=(\xi^{3/2}+\xi^{-3/2})/2\equiv \xi^{3/2}\omega_+/2$.

We wrap the compact D4 brane on $S^4$ which is transverse to the color D4 brane.
Due to the Chern-Simons interaction with RR-field, $U(1)$ gauge field is induced on the D4 brane world volume.
The source of gauge field is interpreted as the end point of fundamental strings. Substituting the equation of motion for gauge field to the
Dirac-Born-Infeld action of D4 brane with the Chern-Simons, we get the Hamiltonian for the compact D4 brane as
\be
{\cal H}_{D4} =\tau_4 \int d\theta \sqrt{\omega_+^{4/3}(\xi^2 + \xi'^2)}\sqrt{D(\theta)^2 +\sin^6 \theta}\, ,
\ee
where $\tau_4 =\frac{1}{2^{2/3}}\mu_4 \Omega_3 g_s^{-1} R^3 U_{KK}$, $D(\theta)=-2+3\cos\theta -\cos^3\theta$ and the prime
denotes the derivative with respect to $\theta$. We assume that the radial coordinate $\xi$  depends only on the polar angle $\theta$ of $S^4$.

\begin{figure}[!ht]
\begin{center}
\includegraphics[angle=0, width=0.42\textwidth]{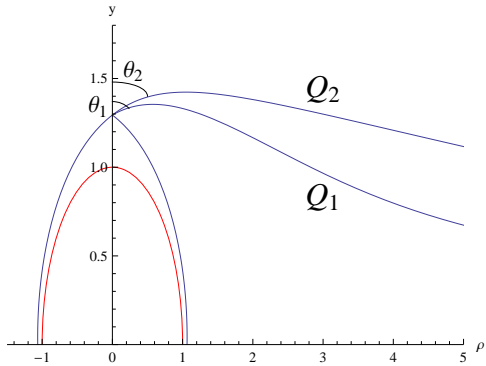}
\caption{(Color online) Embedding of D-branes with $\alpha \ne 0.5$. The asymptotic heights of two branes are the same ($m_1=m_2 =0.1$).  The two branes meet at infinity since $m_1=m_2$. Red curve denotes to the position of $U_{KK}$.} \label{config}
\end{center}
\end{figure}

The fundamental strings out of the compact D4 branes  are attached to two D6 branes,
 and they provide the source of $U(1)$ gauge field on the D6 brane.
 By taking the Legendre transformation for the gauge field, we obtain the Hamiltonian which controls the brane configuration with fixed charge.
\bea
{\cal H}_{D6}= \label{d6h}   \t_6 \int d\r \sqrt{1+\dot{y}^2} \sqrt{\o_+^{4/3}\left(\tilde{Q}^2+\r^4 \o_+^{8/3}\right)} ,
\eea
where  $\t_6 =\frac{1}{4} \mu_6 V_3  \O_2 g_s^{-1} U_{KK}^3$. $\tilde Q$ is dimensionless and related to the number of fundamental strings $Q$ by   $\tilde{Q}=\frac{U_{KK}Q}{2\cdot2^{2/3}\pi\a'\t_6}$.
Baryons are represented by  compact D4 branes, and each of them has $N_c$ fundamental strings attached.
Such configuration of compact D brane plus fundamental strings are called baryon vertex \cite{Witten:1998xy}.
The other ends of fundamental strings  are attached to D6 branes.
Therefore, D6 branes are pulled down and compact D4 brane is pulled up.
As discussed in \cite{Seo:2008qc},
the length of the fundamental strings becomes zero since
the tension of the fundamental strings is always larger than that of D-branes.
  Finally, the position of the cusp of D6 branes should be joined  to that of  the compact D4 brane.
  We consider $Q_1$ fundamental strings attached to one of the D6 branes and $Q_2$ strings to another D6 brane.
  The final configuration is drawn in Fig. (\ref{config}).
We denote the slope at the cusp of each brane as $\dot{y}_c^{(1)}$
and $\dot{y}_c^{(2)}$.
The force at the cusp of D6 branes can be calculated to give
\bea\label{force-d6}
F_{D6}&=&\frac{\partial {\cal H}(Q_1)_{D6}}{\partial U_c} \Bigg|_{\partial}
+\frac{\partial {\cal H}(Q_2)_{D6}}{\partial U_c} \Bigg|_{\partial} \cr\cr
&\equiv& F_{D6}^{(1)}(Q_1)+F_{D6}^{(2)}(Q_2).
\eea
To make the  system stable, following force balancing condition should be satisfied;
\be\label{fbc}
\frac{Q}{N_c} F_{D4} =  F_{D6}^{(1)}(Q_1)+F_{D6}^{(2)}(Q_2),
\ee
where $Q_1=(1-\alpha)Q$ and $Q_2=\alpha Q$ with $0\le\alpha\le 1$, and $F_{D4}$ is the force at the cusp due to the compact D4 brane.
Note that $\alpha=(1-\tilde\alpha)/2$.\par
To find the ground state of our system, we need to consider the energy minimization together with the force balancing condition.
The total energy of our system is
\bea\label{Etot}
E_{tot}&=& \frac{Q}{N_c} {\cal H}_{D4} +{\cal H}_{D6}(Q_1) +{\cal H}_{D6}(Q_2)\, .
\eea
With given $Q$ and $\alpha$ we can calculate the configuration satisfying the force balancing condition.
The energy density per nucleon $E(\rho,\alpha)$ is given by $E_{tot}$ in Eq. (\ref{Etot})
divided by the total baryon number $Q/N_c$.

The symmetry energy is the coefficient of leading term of $\tilde \alpha$ and is a function of $Q$.
Since $Q$ is proportional to the density of the quark or baryon, this way we can calculate the
the symmetry energy as function of density.
  If we further impose minimization of the total energy, we can determine the value of $\alpha$  as a function of $Q$.
For $m_2/m_1\neq 1$,  there exists a transition from a matter with $\alpha=0$ to $\alpha\neq 0$
at a finite value of $Q$. This is identified as a transition from nuclear to strange matter. For very large $Q$, $\alpha$ saturates to $0.5$, as expected. If we take $m_1=m_2$ and if we do not consider isospin violating interactions or electromagnetic interactions, then the ground state of the matter would be always with $\alpha=1/2$.

\begin{figure}[!ht]
\begin{center}
\includegraphics[angle=0, width=0.45\textwidth]{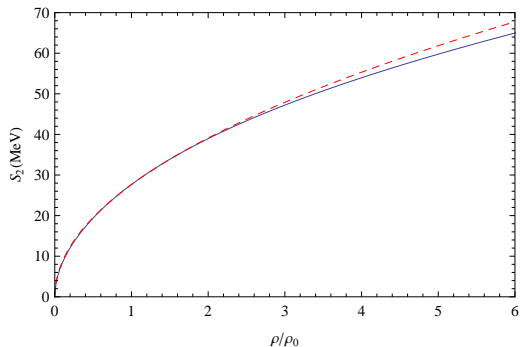}
\caption{(Color online) Solid line is the calculated symmetry energy  as a function of the density.
For illustration purpose, we take  $\lambda=6$ and $M_{KK}=1.04$ GeV.
The  dotted line is for
  $S_2 \sim \rho^{1/2}$.  }\label{symmetryE1}
\end{center}
\end{figure}

The explicit form of symmetry energy per nucleon  can be written as
\be\label{Es}
S_2 =\frac{2\tau_6}{N_B} \int d\r \frac{\sqrt{1+\dot{y}^2}\tilde{Q}^2 \o_+^{10/3} \r^4}{(\tilde{Q}^2 +4 \o_+^{8/3} \r^4)^{3/2}},
\ee
where $y$ is the embedding solution of D6 brane with $\alpha=1/2$.
Notice that
$N_B=Q/N_c$ and so the symmetry energy (\ref{Es}) contains $N_c$ factor.
We need to factor this $N_c$ out for the reason  we discuss in Section \ref{summary}.
Our results are given in Fig. \ref{symmetryE1}.

Note that so far we use $\rho$ for both the coordinate and the density. Hereafter $\rho$ is only for the density.
Although our main point here is the
calculational scheme of the symmetry energy rather than the exact numerical numbers,
it is better to see how the numbers fit to the reality.
To fix the energy scale, we rewrite the energy density per nucleon as
\bea
E(\rho,\alpha)=\frac{\lambda N_c M_{KK}}{2^{2/3}(9\pi)}\frac{\tilde{E}}{\tilde{Q}},   \;\;
\rho = \frac{2\cdot2^{2/3}}{81 (2\pi)^3} \lambda M_{KK}^3\,\tilde{Q}\, ,\label{sR}
\eea
where $\tilde E= E_{tot}/\tau_6$.
 One may determine the values of $\lambda$ and $M_{KK}$ by using the quark mass and meson mass as inputs into two basic
relations: $m_q=\lambda M_{KK}y_\infty/(2^{2/3}9\pi)$ and $M^2_{\eta'}=0.46 M^2_{KK}y_\infty$ with our coordinate choice.
From the  the non-anomalous $\eta^\prime$ mass, $\sim 390$ MeV~\cite{EHS},
and the quark mass, $M_q\sim 41$ MeV, \footnote{As well-known, the quark mass in D4/D6 model
could be different from that in QCD by a constant factor. To obtain the constant we need to compare the scalar two point function
obtained in D4/D6 model
with that in the operator product expansion of QCD, for example see~\cite{PR2}.} we can determine the parameters of the D4/D6 model:
$M_{KK}=1.04$ GeV~\cite{JKS} and $\lambda=6$.
In this case,  $\tilde Q\sim 1.2$ corresponds to the normal nuclear matter density $ \rho_0$.

Now, we lay emphasis on two aspects of our results, which are rather insensitive to the choice of $\lambda$ and $M_{KK}$.
One is the stiffness of the symmetry energy $S_2$ in supra-saturation density regime, and
another is its low density power law  behavior $S_2\sim \rho^{1/2}$.

The power law behavior of $S_2$ in low density can be understood
by calculating  analytically  in a special limit, $m_q \rightarrow \infty$ and $\rho \rightarrow 0$.
In this case, the solution of D6 brane embedding becomes trivial, $\dot{y}=0$ and we can integrate (\ref{Es}) analytically to have
\be
S_2 =\left(\Gamma(\frac{5}{4})\right)^2\sqrt{\frac{ \lambda \rho_0}{2 M_{KK}}} \sqrt{\frac{\rho}{\rho_0}}.
\ee
The current experimental result  of the symmetry energy
can be summarized by a fitting formula
\bea
S_2 (\rho)=c(\rho/\rho_0)^\gamma
\eea
 with $c\simeq31.6$ MeV and $\gamma=0.5-0.7$
in the low density regime, $0.3\rho_0\le \rho\le \rho_0$, see  \cite{sE6} for example.
With  our choice of $\lambda, M_{KK}$, we obtain $\gamma\simeq 0.5$ and $c\simeq 27.7$ MeV from
Eqs. (\ref{Es}) and (\ref{sR}).
Notice that the value of $\gamma$  in our results is rather insensitive to the value of $\lambda$ and $M_{KK}$, while
the value of $c$ depends on them.

One can understand  the  stiffness  based on the property of the
branes: suppose two D6 branes meet with compact D4 at a point
with different polar angles $\theta_1$ and $\theta_2$ as drawn in Fig. \ref{config}.
Here $Q_1$ and $Q_2$ with $Q_1+Q_2=Q$ fixed are the number of the strings
attached to each D6 brane.
The force balance condition is nothing but the minimization of the action with respect to the
position of the contact point.
To balance the pulling force of the compact D4, the position of the contact point is located when each D6 brane balances roughly  half of the pulling force, which is a statement  supported from the numerical analysis.
Since the upward force is proportional to the product of effective tension, $\sim \tau_6 Q$,   times
 the projection factor to the vertical direction, $  \cos \theta$,  we have $Q_1 \cos\theta_1 \sim Q_2\cos\theta_2$. If $\theta_1>\theta_2$, $Q_1>Q_2$. Namely, more strings should be attached to the lower brane in the figure 2.
Therefore, the asymmetry in the number of the attached  string is due to the angle difference.
Notice that each end point of the string provides the `electric' flux
which contributes to the energy of the brane.
 Since  the  flux of charge $Q_i $ is confined in  each brane, the total  energy  is  $ f(Q_1)+f(Q_2)  $, where
 $f(Q) ={\cal H}_{D6}(Q)$. Since $f$ is a monotonically increasing function,
the minimization of the energy with respect to the variation of $Q_1$ requests $Q_1=Q_2=Q/2$.
As Q increases, the effective brane tension increases
and so maintaining the angle difference costs more and more energy.
Furthermore, since $f''(Q)\sim 1/\sqrt{Q}$ is positive,  the symmetry energy, $ Q^2f''(Q/2)/N_B$, is a increasing function.

 The Coulomb repulsion discussed above is, of course,  not the electromagnetic one.
 The local $U(1)$ charge is holographic dual to the the global baryon number of the boundary theory.
However, the repulsion in the  dual bulk theory  means   repulsion in 4 dimension as well.
From the boundary theory point of view, such repulsion is simply due to the presence of the baryon charge.
So the origin of the repulsive nature is mysterious from the boundary point of view.
To understand this, we notice two facts:
The first one is that
the charge carriers are fermions since the charge is introduced by  the D4/D6 fundamental string  end points,  not from the bulk R-charge of type IIA gravity.
The second is that  in the boundary theory the tendency of $N=Z$ by the Pauli principle,
while  in  the holographic dual bulk theory  it is the Coulomb interaction that requests $N=Z$.
Since two origins should be the same,  we may suggest  the Coulomb repulsion as the holographic Pauli principle.
See \cite{Rozali} and \cite{SphereInstanton} for similar observations in a different context.

Finally, we  study the effect of small isospin violation by considering $m_1\neq m_2$.
We find in this case that the symmetry energy is almost the same with the case with isospin invariance for the mass ratios of order one.

\section{Summary and Discussion\label{summary}}
In summary,
we calculated the symmetry energy of dense matter in the D4/D6/D6 model.
To obtain the symmetry energy in nuclear matter with charge symmetry,
 we considered the case with $m_1=m_2$ and found that
 the symmetry energy is increasing with the total charge $Q$,
 showing  a stiff nuclear
symmetry energy. It is  universal in the sense that
the result is independent of the value of $\lambda$ and $M_{KK}$.
We also studied the low density behavior with power  $\gamma$ to be $\sim 1/2$,
which is again independent of the value of $\lambda$ and $M_{KK}$ and is close to the value suggested by experiments, $\gamma=0.5-0.7$.

One subtle point we mentioned in the main text was about the factor $N_c$.
The reason we divided  this factor out is as follows.
In our model, the same flavor quarks form a nucleon; for instance, proton in our model consists of $N_c m_1$ quarks and neutron has $N_c m_2$ quarks in it.
Hence,  the total number difference of quarks is $N_c$ times the number difference of neutrons and protons, resulting in
the overall $N_c$ factor in the symmetry energy. However, in reality, where $N_c=3$,
proton consists of two up quarks and one down quark, and neutron contains one up quark and two down quarks. The total number difference of quarks is the same as
the number difference of neutrons and protons. Therefore, in order to compare our result with the realistic case, we have to divide the symmetry energy (\ref{Es}) by  $N_c$.

We also studied  the effect of  isospin violation by considering $m_1 \neq m_2$.
However, the symmetry energy in this case turned out to be almost the same with the case with isospin invariance for the mass ratios of order one.

Now, we list some generic cautionary remarks.
In holographic approaches, it is not clear how to encode attractive
scalar contributions that are essential to describe nuclear matter.
In addition, for extreme large density limit,
the back-reaction effect from the dense matter is not negligible, and so it should modify our result.

Finally, we comment on a future investigation.
In conventional approaches, for instance see \cite{sE7, Danielewicz:2008cm},
at very low densities $\rho\ll \rho_0$,
the dominant contribution to the symmetry energy is coming from the kinetic energy  which encodes the  Pauli principle.
This is because the kinetic contribution to the symmetry energy is $\sim\rho^{2/3}$,
while the one from interactions starts from $\sim \rho^1$ due to the linear density approximation which works well at very low density.
The origin of the factor $\gamma=2/3$ is the dispersion relation $E\sim p^2$ together with the sharp Fermi surface.
In our case, the fact $\gamma=1/2$  suggests that either the dispersion relation is anomalous like $E\sim p^{3/2}$ or
Fermi surface is fuzzy \cite{sslee, hliu} due to the strong interaction.
This poses an interesting future study.

\acknowledgments

Y.K. and I.J.Shin acknowledge the Max Planck Society(MPG), the Korea Ministry of Education, Science and
Technology(MEST), Gyeongsangbuk-Do and Pohang City for the support of the Independent Junior
Research Group at APCTP.
The work of YS and SJS supported by Mid-career Researcher Program through NRF grant  (No. 2010-0008456)
 and by  NRF grant   through the  CQUeST  with grant number 2005-0049409. SJS was also supported
 by the WCU project (R33-2008-000-10087-0).


\begin{thebibliography}{999}

\bibitem{sE1} P. Danielewicz, R. Lacey, and W. G. Lynch,
Science {\bf 298}, 1592 (2002).
\bibitem{sE2}
 A.W. Steiner, M. Prakash, J. Lattimer, and P. J. Ellis,  Phys. Rept.
{\bf 411}, 325 (2005).
\bibitem{sE3}
B.-A Li, L.-W. Chen and C. M. Ko,
Phys. Rept. {\bf 464}, 113 (2008).

\bibitem{sE5}
  C.~Xu and B.~A.~Li,
  Phys.\ Rev.\  C {\bf 81}, 064612 (2010)
  [arXiv:0910.4803 [nucl-th]].

\bibitem{sE6}
D.V. Shetty and S.J. Yennello, Pramana {\bf 75} 259 (2010).

\bibitem{sE7}
M. Di Toro, V. Baran, M. Colonna, and V. Greco,
 J. Phys. {\bf G 37}, 083101 (2010).

\bibitem{Lee:2010sw}
  H.~K.~Lee, B.~Y.~Park and M.~Rho,
  Phys.\ Rev.\  C {\bf 83}, 025206 (2011)
  [arXiv:1005.0255 [nucl-th]].

\bibitem{XLCYZ}
Z. Xiao, B.-A. Li, L.-W. Chen, G.-C. Yong, and M. Zhang,
Phys. Rev. Lett. {\bf 102}, 062502 (2009).

\bibitem{Che05a} L.W. Chen, C.M. Ko, and B.A. Li, Phys. Rev. Lett. {\bf 94}, 032701  (2005).

\bibitem{Maldacena:1997re}
  J.~M.~Maldacena,
J.M. Maldacena,
Adv.\ Theor.\ Math.\ Phys. {\bf 2}, 231 (1998),
Int.\ J.\ Theor.\ Phys. {\bf 38}, 1113 (1999).


\bibitem{Gubser:1998bc}
  S.~S.~Gubser, I.~R.~Klebanov and A.~M.~Polyakov,
Phys.\ Lett.\ {\bf B428}, 105 (1998).

\bibitem{Witten:1998qj}
  E.~Witten,
Adv.\ Theor.\ Math.\ Phys.\ {\bf 2}, 253 (1998).

\bibitem{Witten:1998xy}
  E.~Witten,
  JHEP {\bf 9807}, 006 (1998).
  [hep-th/9805112].

\bibitem{D4D6_03}
M. Kruczenski, D. Mateos, R. C. Myers and D. J. Winters, JHEP {\bf 0405}, 041 (2004)
[arXiv:hep-th/0311270].
\bibitem{hQCD}
  T.~Sakai and S.~Sugimoto,
  Prog.\ Theor.\ Phys.\  {\bf 113}, 843 (2005)
  [arXiv:hep-th/0412141];
  J.~Erlich, E.~Katz, D.~T.~Son and M.~A.~Stephanov,
  Phys.\ Rev.\ Lett.\  {\bf 95}, 261602 (2005)
  [arXiv:hep-ph/0501128];
  L.~Da Rold and A.~Pomarol,
  Nucl.\ Phys.\  B {\bf 721}, 79 (2005)
  [arXiv:hep-ph/0501218].



\bibitem{Seo:2008qc}
  Y.~Seo and S.-J.~Sin,
  JHEP {\bf 0804},  010 (2008).


\bibitem{KSS2010}
Y. Kim, Y. Seo, and S.-J. Sin,  JHEP {\bf 1003}, 074 (2010).


\bibitem{Rozali}
M. Rozali, H.-H. Shieh, M. V. Raamsdonk, and J. Wu,
JHEP {\bf 0801}, 053 (2008).


\bibitem{SphereInstanton}
  K.~Y.~Kim, S.~J.~Sin and I.~Zahed,
  JHEP {\bf 0809}, 001 (2008)
  [arXiv:0712.1582 [hep-th]].

\bibitem{EHS}
  N.~J.~Evans, S.~D.~H.~Hsu and M.~Schwetz,
  Phys.\ Lett.\  B {\bf 382}, 138 (1996)
  [arXiv:hep-ph/9605267].


\bibitem{PR2}
  L.~Da Rold and A.~Pomarol,
  JHEP {\bf 0601}, 157 (2006)
  [arXiv:hep-ph/0510268];
   A. Cherman, T. D. Cohen and E. S. Werbos,
   {\em The Chiral condensate in holographic models of QCD},
   Phys. Rev. {\bf C79} (2009) 045203.


\bibitem{JKS} K. Jo, Y. Kim, and S.-J. Sin, ``Holographic mesons in D4/D6 model revisited,"
arXiv:1104.2098 [hep-ph].
 \bibitem{sslee}
  S.~S.~Lee,
  Phys.\ Rev.\  D {\bf 79}, 086006 (2009)
  [arXiv:0809.3402 [hep-th]].
  \bibitem{hliu}
  T.~Faulkner, H.~Liu, J.~McGreevy and D.~Vegh,
  arXiv:0907.2694 [hep-th].

\bibitem{Danielewicz:2008cm}
  P.~Danielewicz and J.~Lee,
  Nucl.\ Phys.\  A {\bf 818}, 36 (2009)
  [arXiv:0807.3743 [nucl-th]].

\end{thebibliography}
\end{document}